\documentclass[journal]{IEEEtran}

\usepackage{amsmath}
\usepackage{graphicx}
\usepackage{url}

\begin{document}
\title{Valuating Surface Surveillance Technology for Collaborative Multiple-Spot Control of Airport Departure Operations}

\author{Pierrick~Burgain,~
        Sang~Hyun~Kim,~
        and~Eric~Feron~
\thanks{P. Burgain is with Capital One, Atlanta, GA 30348-5474 USA (e-mail: pburgain@gmail.com).}
\thanks{S.H. Kim is with the School of Aerospace Engineering, Georgia Institute of Technology, Atlanta, GA 30332 USA (e-mail: sanghyun.kim@gatech.edu).}
\thanks{E. Feron is with the School of Aerospace Engineering, Georgia Institute of Technology, Atlanta, GA 30332 USA, and with Ecole Nationale de l'Aviation Civile, 31000 Toulouse, France (e-mail: feron@gatech.edu).}
}


\maketitle

\begin{abstract}
Airport departure operations are a source of airline delays and passenger frustration. Excessive surface traffic is a cause of increased controller and pilot workload. It is also a source of increased emissions and delays, and does not yield improved runway throughput. Leveraging the extensive past research on airport departure management, this paper explores the environmental and safety benefits that improved surveillance technologies can bring in the context of gate- or spot-release strategies. The paper shows that improved surveillance technologies can yield 4\% to 6\% reduction of aircraft on taxiway, and therefore emissions, in addition to the savings currently observed by implementing threshold starategies under evaluation at Boston Logan Airport and other busy airports during congested periods. These calculated benefits contrast sharply with our previous work, which relied on simplified airport ramp areas with a single departure spot, and where fewer environmental and economic benefits of advanced surface surveillance systems could be established. Our work is illustrated by its application to New-York LaGuardia and Seattle Tacoma airports.
\end{abstract}


\section{Introduction: Lean Airport Operations Mean Fewer Aircraft on the Taxiways}

\subsection{Air Traffic Growth and Airport Congestion}
\IEEEPARstart{T}{he} U.S. National Airspace System (NAS) is expected to grow about 2.4\% per year over the next 20 years and accommodate around 1.6 times 2008's traffic level by 2028 \cite{greenair2010,nextGen,Ky2006,Arbuckle2006}. The anticipated growth of air traffic is expected to bring additional load to an already congested system \cite{Joint2004, Thanh2006}. Even though the development of smaller regional airports is expected, it is predicted that major airports will keep running at full capacity \cite{nextGen}. In some cases, airports will not be able to expand their capacity sufficiently to meet the increasing demand. Airports like New-York LaGuardia are physically restrained by the lack of space for new runways or ramps. Other airports are not able to grow physically because of significant opposition from local communities. Therefore, such airports are bound to be congested.

\subsection{Environmental Impact of Airports}
The contribution of aviation to CO$_{2}$ and NOx emissions around airports is expected to increase significantly by 2025 and beyond \cite{U.S.1999,Intergovernmental1999}. Hence, environmental impacts are expected to be a fundamental constraint on air transportation growth \cite{Waitz2004}. Indeed, concerns over pollution have forced governmental, environmental, and regulatory agencies to start implementing emissions abatement procedures at certain airports, such as LaGuardia Airport \cite{Boeing}. In 2011, European Commission decided to regulate CO$_{2}$ emissions at the average 2004-2006 levels \cite{eucap}. This legislation will be applied to all the flights arriving at and departing from European airports. In the United States, Section 231 of the Clean Air Act gives the Environmental Protection Agency (EPA) the authority to regulate aircraft emissions and to adopt emissions standards for U.S.-flagged aircraft \cite{United2003}. Additionally, many efforts are being conducted towards emissions-reduction technologies and concepts, such as electric taxi and new operational procedures. The latter is expected to provide the greatest near-term benefits \cite{Waitz2004}. On the ground, the level of environmental nuisance generated by chemicals and noise can be directly tied to the number of aircraft whose engines are running at any given time. These aircraft are typically those in the taxi phase. Thus, for a given level of airport service, e.g., target hourly number of operations, the fewer aircraft taxiing on the airport surface at any time, the lower the environmental impact.

\subsection{Current Initiatives for Improving Departure Operations}
To tackle the environmental issue, the NextGen concept of operations \cite{nextGen} encourages research in surface traffic operations aimed at lowering emissions and improving surface traffic planning. Likewise, according to \cite{Eurocontrol2011,cdmmanual}, EUROCONTROL is currently fine-tuning the Airport Collaborative Decision Making Departure Manager (CDM DMAN) concept of operations and is preparing the necessary implementation guidelines. DMAN incorporates Collaborative Decision-Making (CDM) as a tool for managing departure operations. According to \cite{dman}, DMAN keeps the number of aircraft on the taxiway at an optimal level and keeps the taxiways open for other traffic without blocking stands for arrivals, reduces controller workload, improves punctuality and predictability, facilitates co-operation between aerodrome ATC, airlines and airport operators, enhances CFMU [i.e. Central Flow Management Unit slot-revisions] and slot compliance, and exploits the departure capacity of the respective runway. Recent work by Bohme et al. \cite{bohme2007coordinated} proposed a coordination of Arrival and Departure Manager (AMAN and DMAN) in order to increase the efficiency, punctuality, and predictability of the departure operations. As of today, Airport CDM has been implemented at some major European airports including Munich Airport, Brussels Airport, Frankfurt Airport and Paris-Charles de Gaulle, and it is scheduled to expand to more European airports \cite{Eurocontrol2011}. In the U.S., the Collaborative Departure Queue Management was evaluated at Memphis International Airport \cite{brinton2011collaborative}, and the Surface Congestion Management Technique was field-tested at New York JFK Airport \cite{nakahara2011analysis}. Also, NASA developed the Spot and Runway Departure Advisor tool to help controllers manage airport surface operations and evaluated the tool in a human-in-the-loop simulation for Dallas/Fort Worth International Airport \cite{jung2011performance}.

\subsection{Analytic Research Efforts}
\begin{figure}[!t]
\centering
\includegraphics[scale = .30]{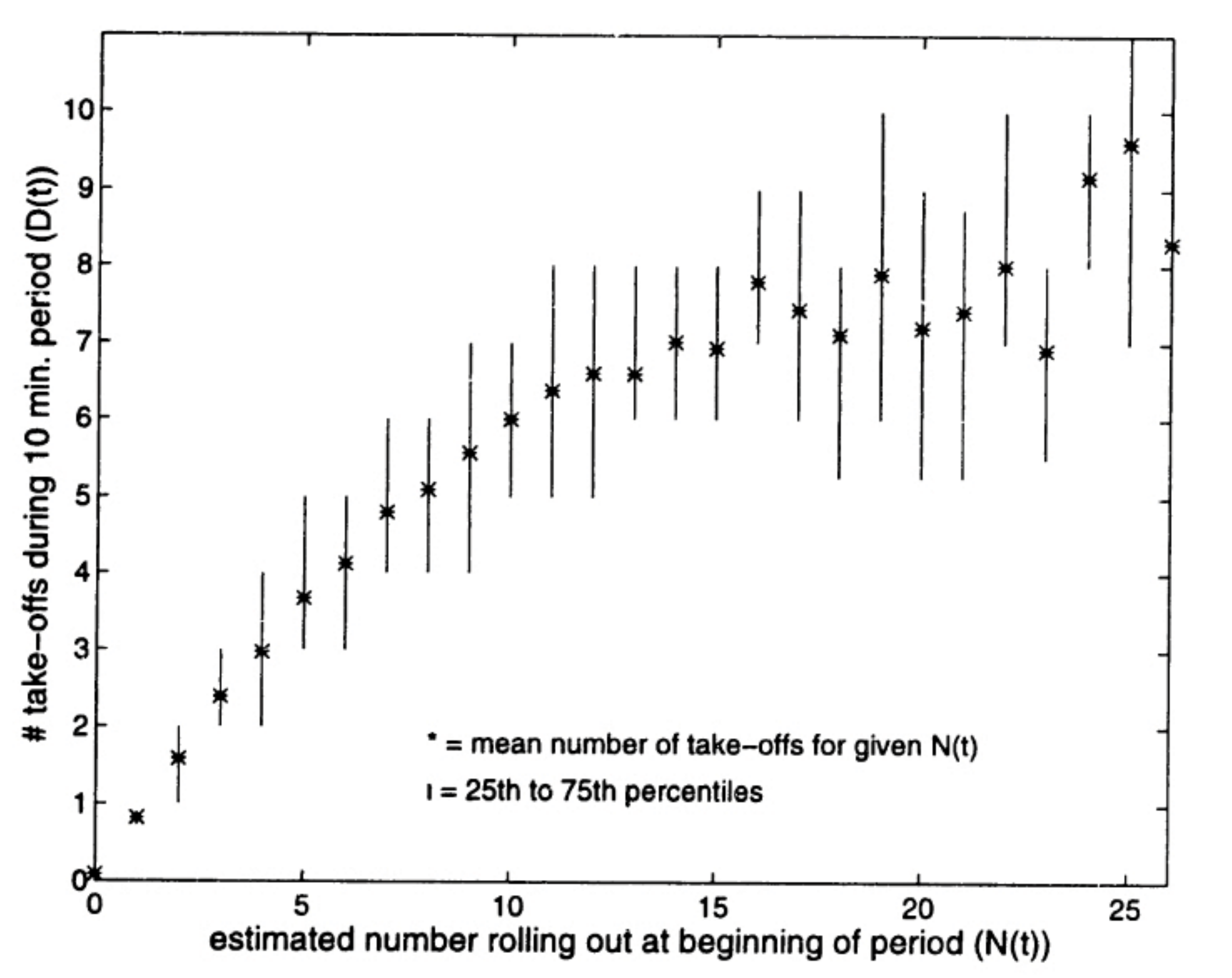}
\caption{Boston Logan Airport Throughput as a Function of Surface Congestion~\cite[p. 82]{Shu:95}: N(t) is the number of taxi-out aircraft at time t.}
\label{shumskypix}
\end{figure}

The fundamental observation supporting most recent research efforts is the existence of a close relationship between the number, or density, of aircraft buffered between the gate and the runway, and the runway throughput. First observed experimentally by Shumsky~\cite{Shu:95}, when considering Boston Logan airport, the runway throughput grows with the number of aircraft buffered between the gates and the runway; however, the throughput saturates past a given level of surface congestion, as shown in Fig.~\ref{shumskypix}. An asterix indicates the mean number of take-offs. Each vertical bar is the range from first to third quartile. Note how airport throughput tends to saturate when the number of aircraft taxiing-out exceed about 15. 

From this observation, a number of steps followed: In~\cite{Feron1997}, Feron {\em et al.} discuss the creation of a virtual queue to control aircraft access to the taxiway system, while respecting the first-come, first-serve rule that dominates air traffic management. Pujet {\em et al.}~\cite{log99} develop a detailed queuing model of airport departure operations and introduce a simple thresholding scheme to regulate departures: Pushbacks are allowed only to the extent that the number of aircraft present on the taxiway system (the buffer), does not exceed a given threshold. Carr {\em et al.}~\cite{Carr2002} describe an approach for modeling and controlling queueing dynamic under severe flow restrictions and Idris et al. \cite{Idris2002} develop a queueing model for the purpose of estimating taxi-out times. Recent developments include~\cite{Burgain2009}, which details the potential benefits of intra-airline slot-swapping inside the virtual departure queue. Finally, two notable efforts have led to field implementations of virtual queueing concepts. In~\cite{SKB:11}, Simaiakis {\em et al.} describe the experimental implementation of a congestion control scheme by means of windowing derived from that proposed in~\cite{log99} and they report significant actual fuel savings and emission reductions: According to~\cite{SKB:11}, the fuel savings are of the same order of magnitude as those generated by Continuous Descent Approaches \cite{cda}. The project CDM@CDG (see \url{http://www.euro-cdm.org/library/airports/cdg/}) has initiated the implementation of a departure manager for Charles de Gaulle Airport, whose purpose is to reduce airport surface congestion due to departing traffic. The virtual queueing effectively created is leveraged by CDG's major airlines to perform departure swaps of the kind investigated in~\cite{Burgain2009} through the implementation and evaluation of SESAR's Dflex program \cite{dflex}.

\subsection{New Surface Surveillance Information}
The gate- and spot-release control efforts described above can be easily implemented by means of existing technology. Indeed, the only information needed for implementation is already available in various forms, including ACARS~(Aircraft Communications Addressing and Reporting System), for which commercial decoding systems are now available. What motivates this paper and its predecessor~\cite{burgain2011}, however, is that airports become progressively equipped with modern, digital surface surveillance technologies, such as the Airport Surface Detection Equipment, model X (ASDE-X) and Advanced Surface Movement Guidance \& Control System (A-SMGCS). With such systems, accurate aircraft ground position information becomes more easily available in real-time~\cite{ICAO2004,Eurocontrol2008,Emma2007,Aviation2009,Besada2004,Dimitropoulos2007}. Primarily designed for improved surface operations safety, the impact of these systems on the reduction of runway incursion incidents and conflicts has been the focus of several studies~\cite{Young2000,Singh2004,Schofield2008}.

Other studies, however, also focus on the impact of advanced surface surveillance on airport efficiency, for example to precisely control taxiing aircraft and increase the efficiency of active runway crossings~\cite{Cheng2001}. Early experiments show significant operational improvements enabled by airport surface surveillance technologies. Howell et al.~\cite{Howell2004}, for example, directly measure the impact of surveillance data sharing on surface operations at Memphis International Airport and at Detroit Metropolitan Wayne County Airport. They show that surface surveillance data made available to ground controllers directly lead to shorter taxi times. At Memphis airport, average taxi time is reduced by 6.6 percent during Visual Approach conditions (visibility greater than five miles and ceiling greater than 5000 feet) , and by 17.5 percent during Instrument Approach conditions. In another field study~\cite{Howell2005}, Howell et al. take advantage of a surface surveillance outage to examine its impact on airline operations. They measure changes in taxi-out times, queue lengths, and departure rates before, during, and after the outage. They find that, for similar levels of airport surface queues, surface surveillance decreases taxi-out times. Furthermore, recent work investigates the integration of surface surveillance for aircraft arrivals in a collaborative environment~\cite{Andersson2003,Jung2005,Hughes2006}.

\subsection{Contributions of this Paper}
The prior work described above leaves open, however, the analytical evaluation of the impact of improved surveillance technologies on gate- or spot-release strategies. We propose to perform this evaluation by revisiting the gate- or spot- release strategies introduced in \cite{log99} and asking whether the performance of such gate- or spot- release strategies can be improved using vastly improved data on aircraft position. Intuitively, this should be the case: A cluster of five departing aircraft near the runway threshold should prompt decisions that differ from those required if the same cluster of five departing aircraft has just left their gate or spot. The two situations, however, are considered to be equivalent under the policies discussed in~\cite{log99,SKB:11}. In addition, some of the surveillance systems are not available at some airports, or the data they provide cannot be easily used. Gauging the benefits of a particular surveillance technology can help the airports decide on whether it is worth purchasing the system altogether. Because ASDE-X reliably covers only the Airport Movement Area (AMA) such as runways and taxiways, the remaining sections of this paper therefore aim at exploring the potential benefits of aircraft position information on spot-release strategies: Section 2 describes a modeling approach of busy airports similar to that introduced in \cite{burgain2011} by means of finite-state Markov Decision Processes. The improvement of this paper over \cite{burgain2011} is the existence of multiple ramp areas. Unlike \cite{burgain2011}, in which the airport model is simplified by assuming the existence of a single ramp area, the airport model in this paper is more complex and closer to actual airports including multiple ramp areas. Section 3 then discusses optimal spot-release strategies and discusses the efficiency gains that may be expected from using surface surveillance information.

\section{Modeling Busy Airports by Means of Markov Decision Processes}
To clarify the impact of added surface information on spot-release strategies, we study feedback control laws under various information scenarios. We define surface information in terms of aircraft ground position on the AMA, ramp access to the taxiway system, and runway queue length. Similar to \cite{burgain2011}, a stochastic model of taxi departure operations is developed by means of Markov Decision Processes (MDP) and Partially Observable Markov Decision Processes (POMDPs). However, the stochastic model simulates more complex taxiway systems composed of multiple ramp areas. All numerical computations of model calibration are done using a model of New York LaGuardia Airport and its operations. Also, the model is expanded to Seattle Tacoma Airport.

\subsection{Airport Modeling} 
\label{s:1rModel}
Airport operations are modeled as a Markov Decision Process. The proposed stochastic model emulates departure surface operations when:
 \begin{itemize}
 \item Exact aircraft positions on the AMA are available,
 \item Aircraft trajectories are subject to uncertainties.
 \end{itemize}
Markov Decision Processes are attractive because numerical procedures are well-identified to compute steady state optimal control policies, using linear programming \cite{hillier1990introduction}. This is unlike the models discussed in~\cite{log99}, whose resolution is finer, but which can be used only to simulate elementary control laws such as windowing schemes. The airport surface is discretized by representing it as a finite number of ``boxes'', within which aircraft may be found. Thus, the number of aircraft locations is finite, and depends upon a spatial sampling of the taxiway system. At each time step, aircraft may move to the next available spatial sample or stay in place. 

\subsubsection{Model Description}
When a taxi clearance is issued by ground controllers, an aircraft enters the AMA through a spot. Aircraft motion along the taxiway is described by state transition maps that describe the probabilities for aircraft to move forward or stay in place. When aircraft arrive at the runway threshold, they enter a limited capacity buffer directly servicing the runway, and the aircraft order is maintained on the runway queue. The take-off clearance process is then simulated as a steady state stochastic process using the sum of two Bernoulli variables. This sum provides the means to calibrate not only the average, but also the standard deviation of the take-off rate. The approach borrows from previous models~\cite{log99}. The uncertainty related to the take-off time illustrates the limited prediction capabilities that agents issuing ground clearances have regarding the exact take-off clearance time.

\subsubsection{Surface States Coding}
\begin{figure}[!t]
\centering
\includegraphics[width=3.5in]{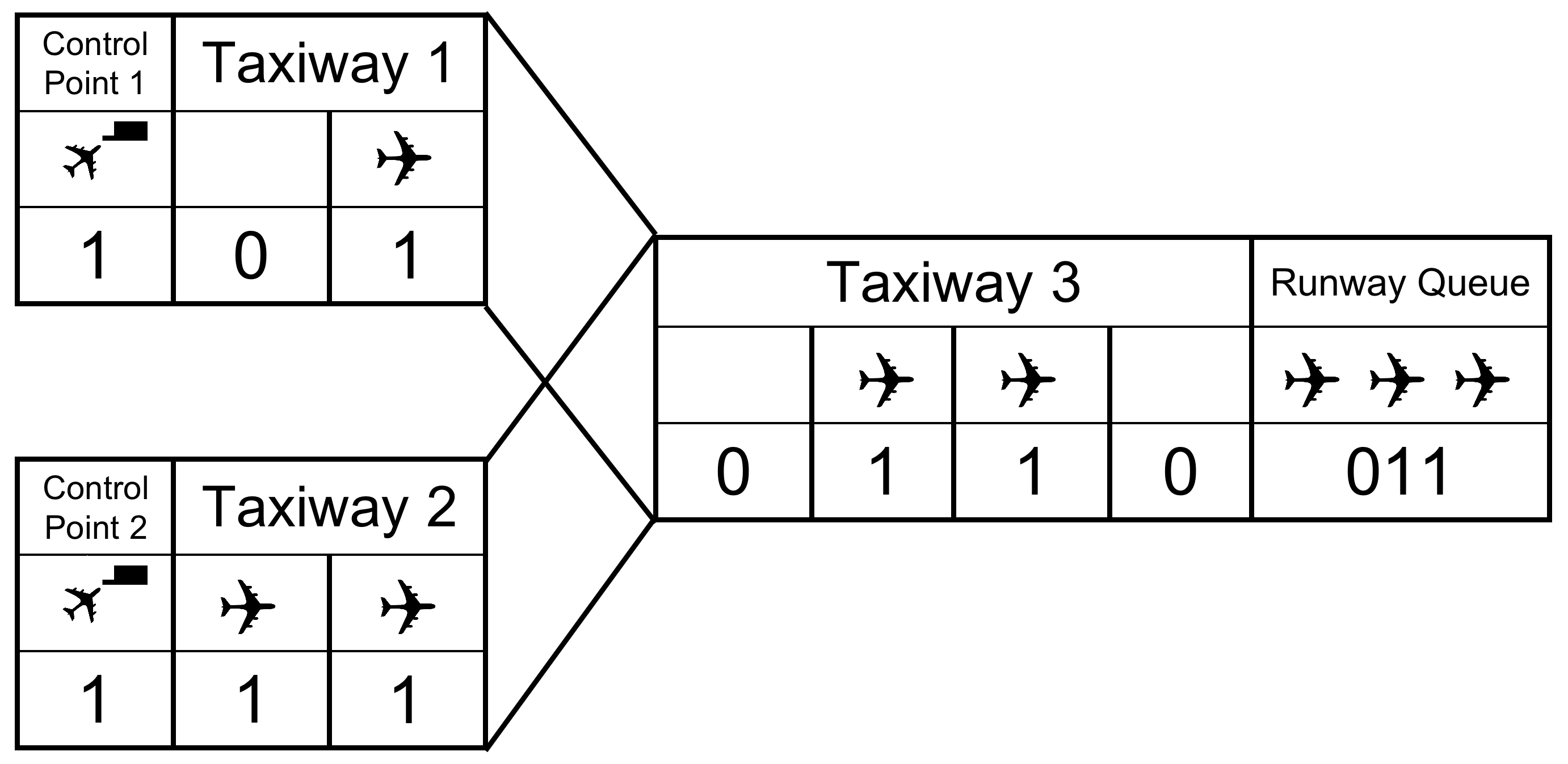}
\caption{State Space Model of Typical Airport by Means of Markov Decision Processes}
\label{modeling}
\end{figure}

Each state is represented by a binary vector composed of three fields: the control points, the taxiways, and the runway queues, as illustrated in Fig.~\ref{modeling}. In the first field, a control point represents the entry point of a taxiway, such as a spot. When a taxi clearance is issued, one of the control points is switched from 0 to 1 to indicate that an aircraft at the corresponding spot location has been cleared to taxi toward the runway. We focus on busy hours of the airport, and it is assumed that there are always aircraft ready to taxi from every ramp area. However, an airport such as Dallas/Fort Worth Airport with over 50 spots does not need aircraft waiting at every spot to be congested. We simplified the analysis by aggregating all spots in a ramp area and assuming that the aggregate spots always have an aircraft ready to go. Note that we could extend the model to include individual and non-congested spots, but We do only "coarse, analytic evaluations". A full study would require fast-time, higher-detailed simulation efforts that go beyond the scope of this paper. The second field is the taxiways, which are directly connected to the control points. The taxiways are spatially sampled, with only one aircraft allowed per spatial sample. The state of the taxiways is represented by a binary vector whose size is equal to the number of spatial samples. The vector's elements are set to one when the corresponding spatial sample is occupied by an aircraft, and zero otherwise. The runway threshold queue state is expressed as a binary number representing the number of aircraft in the queue. For instance, if there are 3 aircraft queueing at the runway threshold, the state of the queue is given by the binary vector 011. The entire state vector is then obtained by concatenating the binary fields of the control points, the taxiway system, and the runway queues. Finally, the overall binary vector is converted to a decimal number, which is its state identification number. For instance, Fig.~\ref{modeling} represents the state vector 1101110110011.

\subsubsection{Indices and Notations}
The state space and the state index space are linked by a bijective index function. In the rest of this paper, the notation $i$ refers to the index of a state vector. The notation $i[s]$ refers the $s^{\rm{th}}$ component of the state vector $i \in S$.

\subsubsection{Model Parameters}
The system is entirely specified by the following parameters:

\begin{itemize}
\item $L_s$ : The taxiway length represented by one spatial sample,
\item $T_s$ : The sampling time,
\item $N$: The number of spatial samples,
\item $m$: The probability of moving forward at the next time step,
\item $c_1$ and $c_2$: The probability of receiving a take-off clearance for an aircraft at the runway threshold is determined by two Bernoulli variables with parameters $c_1$ and $c_2$, where a Bernoulli variable can take value 1 with probability $c$ and value 0 with probability $1-c$,
\item $B$: The maximum capacity of the runway threshold aircraft buffer.
\end{itemize}

Departure operations are modeled as a Markov Decision Process. Thus, they are entirely defined by the probabilities of transition from a state $i$ to another state $j$, knowing that  the decision to send an aircraft on the taxiway is $k$ (e.g., $k=1$ corresponds to the decision of sending an aircraft from ramp 1, and $k=3$ corresponds to the decision of sending aircraft from ramp 1 and ramp 2). These probabilities are the model transition probabilities, and they are noted $P_{j|ik}$. These probabilities are evaluated from the parameters described above. To give some idea of the model complexity, a typical airport may contain about 220,000 nonzero transition probabilities.

\subsubsection{Markov Decision Process: States and Transition Probabilities} \label{s:transitions}
The transition probabilities are generated by enumerating all possible simultaneous sub-transitions that lead to a feasible state. Sub-transitions are defined as atomic transitions that happen during the same time step. The process by which these transition probabilities are generated is tedious and the reader is invited to refer to~\cite{Bur:10} for more details.

\subsection{Model Calibration Procedure} \label{s:calibration}
\begin{figure}[!t]
\centering
\includegraphics[scale = .3]{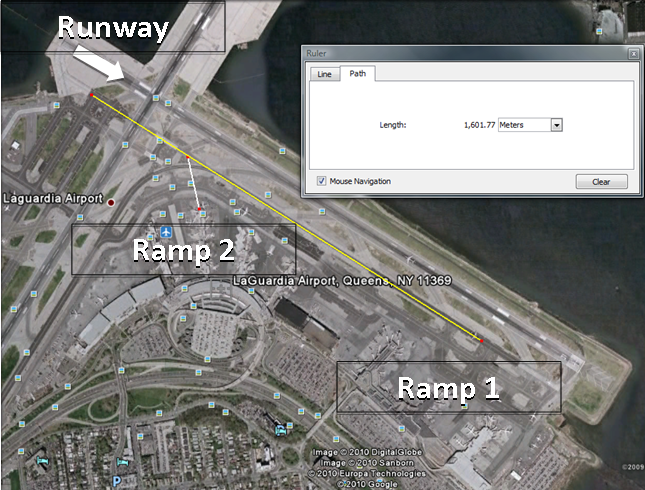}
\caption{LaGuardia Airport in Most Common Configuration.}
\label{f:LGQuickAccess}
\end{figure}

\begin{figure}[!t]
\centering
\includegraphics[scale = .3]{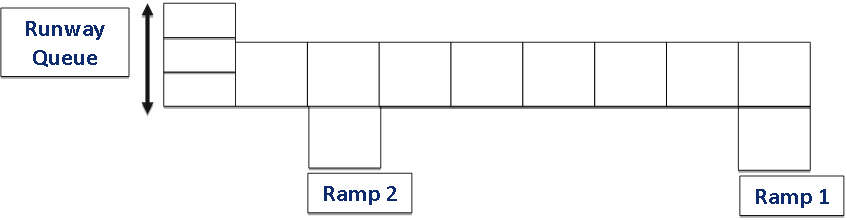}
\caption{A Model of LaGuardia as Markov Decision Process: Terminal C and D are dealt with a single large ramp area (Ramp 1), and Terminal A and B consist of Ramp 2.}
\label{f:2ndModel}
\end{figure}

The calibration of the model is based on the analysis of selected Aviation System Performance Metrics (ASPM) data, as well as direct observations of airport satellite pictures. ASPM contains flight data such as flight number, departure/arrival airports, departure/arrival times, departure/arrival delays, etc. The following quantities are defined: $L_s$ corresponds to observations of physical distances between taxiing aircraft. $T_s$ is defined as the shortest characteristic time of the different phenomena captured by the model. The variables $N$, the number of spatial samples of the taxiway, and $m$, the probability of moving forward when unencumbered, are calibrated using taxi statistics derived from ASPM data. Finally, $c_1$, $c_2$, which define the take-off probabilities, and $B$, the runway buffer size, are calibrated using take-off statistics coupled with estimates of the number of taxiing aircraft. The calibration procedure is now applied to New York LaGuardia airport, shown in Fig.~\ref{f:LGQuickAccess}, and its operations for the year 2006. Owing to the presence of two main terminals, the airport is represented using the Markov Decision Process illustrated in Fig.~\ref{f:2ndModel}.

The following quantities are identified:

\subsubsection{Sampling Time}
 The temporal resolution of the ASPM data is one minute. Our model sampling frequency was set to match the sampling rate of the data against which it is calibrated. Thus, $T_s$ is set to one minute.

\subsubsection{Departure Capacity}
 Heavy traffic surface operations are used to evaluate the departure capacity and to calibrate the take-off clearance variables $c_1$ and $c_2$. Heavy traffic corresponds to the number of aircraft for which the average number of take-off per minute saturates. In the case of LaGuardia Airport, heavy traffic is achieved when 14 or more aircraft are taxiing toward the runway. Data show that the airport take-off rate has a mean of 0.605 aircraft per minute and a standard deviation of 0.578 aircraft per minute when the taxiway system is saturated. Details on the departure capacity of LaGuardia Airport are given in \cite{burgain2011}. The departure capacity is known to be dependent on arrival rate \cite{SKB:11}, but the number of taxi-out aircraft is dominating. Hence, in this paper, arrival rate is not considered for the calibration of departure capacity for the simplicity. The take-off clearances are modeled using the sum of two Bernoulli variables $c_1$ and $c_2$ equal to 0.5140, and 0.0929, respectively (variables following a Bernoulli distribution of parameter $p$ are equal to 1 with probability $p$ and 0 with probability $1-p$). The sum of the two random variables is evaluated at every minute and determines how many aircraft take off. 
The value of these two parameters was determined by solving the following system of equations:
\begin{eqnarray} \label{e:c1c2equation}
  \mbox{Average} = c_1 + c_2 = 0.605 \mbox{ aircraft/minute}
  \end{eqnarray}
  \begin{eqnarray}
\mbox{Std Deviation} &=& \sqrt{c_1\cdot (1-c_1) + c_2\cdot (1-c_2)} \nonumber \\
&=& 0.578 \mbox{ aircraft/minute}
\end{eqnarray}

\subsubsection{Taxiways}
\begin{figure}[!t]
\centering
\includegraphics[width=3.5in]{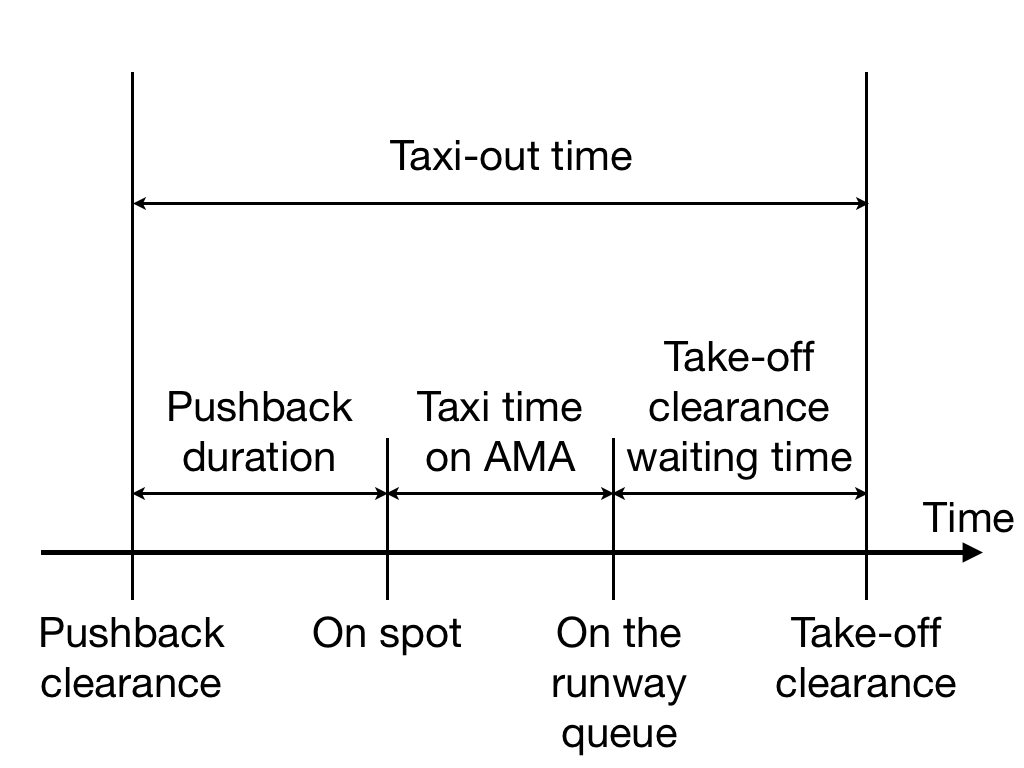}
\caption{Taxi-out Time}
\label{taxitime}
\end{figure}

Once the departure rate variables are calibrated, the taxiway variables $N$ and $m$ are calibrated to reproduce light-traffic unimpeded taxi-time average and standard deviation for aircraft pushing back from each ramp area. The standard deviation and average of light-traffic taxi times are evaluated using the ASPM database. The taxi-out time is defined as the time between push-back and wheels-off and includes pushback, taxi, and waiting for take-off clearance times. All the times and events that occur during the departure process are shown in Fig. \ref{taxitime}. Only the taxi-out time is observable from the ASPM data, because the ASPM data do not record any event between the push-back time and the take-off time.
Therefore average taxi times on the AMA are computed by subtracting average pushback durations and average take-off clearance times from average taxi-out times. Likewise, taxi-time variances are computed by subtracting pushback duration variances and take-off clearance time variance from taxi-out time variances. The location of a ramp area is identified using the unimpeded taxi-out time, which is a quantity in the ASPM data that is estimated from a regression equation and is not physically observable: aircraft taxiing from the same or nearby ramp areas to the same runway are very likely to have similar unimpeded taxi-out times.
\begin{itemize}
\item Unimpeded taxi-out times: These taxi-out times were computed by considering taxi-out times when surrounding traffic is low. For Ramp 1, for instance, the average is 13.56 minutes and the standard deviation is 2.00 minutes. Note that the taxi-out times depend on the location of the gate and the taxi route, so the unimpeded taxi-out times from the same ramp are not necessarily the same.

\item Pushbacks: Average duration of pushback was evaluated by Delcaire and Feron \cite{Delcaire1997} at 2 minutes. Based on the data collected in their report, it is fair to estimate the standard deviation of pushback duration at 80 seconds, or 1.33 minutes.

\item Take-off clearance: Taxi-out times include waiting times for take-off clearance at the runway threshold. However, the model calibration should not include the variation caused by this waiting time. In this model, the average waiting time for one aircraft at the runway threshold before clearance is $1/0.605 = 1.65$ minutes, and the standard deviation is $1.04$ minutes.

\item Taxi time on the AMA: According to the above discussion, the taxi time on the AMA from ramp 1 has a standard deviation of $\sqrt{2.00^{2}-1.04^{2}-1.33^2}=1.07$ minutes and an average of $13.56-1.65-2 = 9.91$ minutes. A similar process for Ramp 2 yields an average taxi time on the AMA from Ramp 2 equal to 6.4 minutes.

The probability $m$  of moving forward on the taxiway system, and the number of steps $N$ from each ramp to the runway threshold, were calibrated to match the average and standard deviation of taxi times in light traffic under nominal conditions. For Ramp 1, $N$ and $m$ solve the following system of equations.
\begin{eqnarray}
\mbox{Average} = \frac{N}{m}\cdot T_s  = 9.91 \mbox{ minutes}
\end{eqnarray}
$$ \mbox{Standard Deviation} = $$
 \begin{eqnarray}
\frac{N}{m}\cdot \sqrt{\frac{1-m}{N}} \cdot T_s  = 1.07 \mbox{ minutes.}
\end{eqnarray}

Thus,
$$N = 8.88 \approx 9 \mbox{ steps}$$
 $$m = 0.90 \approx \frac{9}{\mbox{Average}} = 0.9084.$$
For Ramp 2, we find $N=3$ and $m$ remains the same.
\item Runway buffer capacity $B$:
The aircraft buffer at the runway threshold simulates aircraft that stand close to each other in order to ensure a high utilization rate. The buffer capacity must be as small as possible to limit the size of the state space over which optimal policies are computed. However, the buffer capacity needs to be large enough to allow ground controllers to absorb uncertainties in take-off clearance time and taxi time. The standard deviation yielded by the sum of these two times for a single aircraft is $\sqrt{1.07^{2}+1.04^{2}}=1.49$ minutes.

The buffer was calibrated to be able, when fully loaded, to supply aircraft for a time close to 3 times this standard deviation, i.e. $4.47$ minutes. Thus,  the buffer size was approximated to provide enough aircraft to cover at least $4.47$ minutes, which is $4.47/0.605 = 7.39 \approx 7$ take-off clearances. The capacity was set to 7 aircraft and the buffer was coded using 3 bits, as illustrated in Fig.~\ref{modeling}.

\item Physical distance between aircraft $L_s$:
A 200-meter separation between taxiing aircraft was suggested in previous work on taxi operations \cite{Balakrishnan2007,Visser2003}. Hence, that number was adopted here as well.\\
\end{itemize}
The calibration values for the system parameters are summarized in Table \ref{table:calibration}.

\begin{table}[!t]
  \caption{Calibration Values}
  \begin{center}
  \begin{tabular}{c|c}
  \hline
  \hline
  \textbf{Calibration Variables} & \textbf{Values}\\
  \hline
  $L_{s}$ & 200 meters\\
  $T_{s}$ & 60 seconds\\
  $N$ & 9 (Ramp 1)\\
  & 3 (Ramp 2)\\
  $m$ & 0.9084\\
  $c_1$ & 0.5140\\
  $c_2$ & 0.0929\\
  $B$ & 7\\
  \hline
  \hline
\end{tabular}
\end{center}
\label{table:calibration}
\end{table}

\subsubsection{Model Validation}
\begin{figure}[!t]
\centering
\includegraphics[width=3.5in]{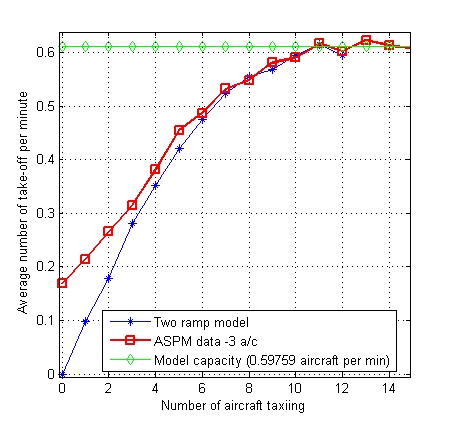}
\caption{LaGuardia Throughput as a Function of the Number of Taxiing Aircraft, from the Two Ramp Model and ASPM Data. 
The ASPM curve is shifted by 3 aircraft to isolate taxiway operations starting at ramp exit control points, for utilization rates above 30\%.}
\label{loadGraph22}
\end{figure}

Using ASPM data, LaGuardia airport average throughput rate is expressed as a function of the number of taxiing aircraft. The graph provided in Fig.~\ref{loadGraph22} shows the airport throughput as a function of the number of taxiing aircraft, and yields the average take-off rate. Fig.~\ref{loadGraph22}  also shows the throughput as a function of the number of taxiing aircraft for the stochastic model. The model behaves similarly to the airport, and faithfully reproduces the queueing and stochastic nature of departure operations. Indeed, when the number of taxiing aircraft reaches 11, the model saturates, and yields a maximum take-off rate distribution averaging 0.598 aircraft per minute, with a standard deviation of 0.585 aircraft per minute. These numbers are similar to the average (0.605) and the standard deviation (0.578) of the observed take-off rate at LaGuardia, when the taxiway is saturated by departing aircraft. The saturation level of the model take-off rate is reached at a lower number of taxiing aircraft than for the ASPM data because the model accounts for operations on the taxiway only from the ramp control points. By contrast, the ASPM data includes all aircraft on the ground starting at pushback. The ASPM data does not provide aircraft position, therefore it is not possible to distinguish aircraft still pushing back at the ramp from aircraft which are at the ramp exit control points. To isolate taxiway operations starting at the control points from the rest of the ramp operations in the ASPM data, the ASPM curve has been shifted to match the saturation level of both curves. For runway utilization rates above 30\% of interest in this paper, the shift efficiently isolates taxiway operations starting at the control points in the ASPM data, as illustrated in Fig.~\ref{loadGraph22}. Note that the two-ramp model of LaGuardia airport reproduces the ASPM data better than the one-ramp model discussed in \cite{burgain2011}.  


\section{Quantitative Impact of Full-State Information: Optimal Control of Airports Represented as MDPs}
To understand and evaluate the impact of aircraft position information on departure operations, an approach based on the optimization of Markov Decision Processes (MDP) and Partially Observable Markov Decision Processes (POMDP) was developed. This approach is applied to two distinct state-based policies and one benchmark policy:
\begin{itemize}
\item The first policy, named ``optimal full state feedback'' assumes that the state of the airport surface (in terms of aircraft location) is fully available.
\item The second policy, named ``estimated state feedback'' assumes that the only part of the state of the surface is known, and the most likely state is used for feedback.
\item The benchmark policy, named ``threshold policy'' tries to maintain the number of taxi-out aircraft under a certain threshold, and this policy is a variant on that used in prior analytical and experimental works~\cite{log99,SKB:11}.
\end{itemize}

\subsection{Approach} \label{theappraoch}
The objective of this section is to evaluate how the level of information available on aircraft position affects potential taxi-time reductions, for a given rate of runway utilization, and within a collaborative framework enabling the fine tuning of taxi clearances, when aircraft exit the ramp area.

\subsubsection{Assumptions}
Ground controllers operate as optimally as allowed by existing technology: they know the behavior of the system, and given the level of information available, they understand the best policy for maximizing the departure runway utilization rate while controlling aircraft to minimize taxi times. It is assumed that there is enough departure demand for FAA ground controllers to always have an aircraft waiting to be cleared for taxi at both ramps, since this configuration corresponds to peak demand times. The aircraft is either cleared for push-back, if it pushes directly on the movement area, or cleared for taxi, if it has already pushed back on the ramp area, and is waiting at a control point to enter the movement area.

\subsubsection{Optimal Pushback Policies}
Each state has a cost, and an optimal clearance policy is the set of spot-release decisions that minimizes the expected cost, averaged over an infinite time horizon. With the assumption of an infinite time horizon, the expected averaged cost is easily calculated using steady state probabilities. 

\subsubsection{Trade-offs and Cost Structure}
For each time instant, each state $i$ is given a cost $C_{i}$ that reflects its desirability. This cost is a weighted sum of the number of taxiing aircraft  $N_{ac}(i)$ and a cost attributed to the non-utilization of the runway $\delta_{r}(i)$ multiplied by a constant $ \beta $. The variable $\delta_{r}(i)$ is equal to 1 if there is no aircraft in the runway buffer and to 0 if there is at least 1 aircraft.
For every state $i$, the cost $C_{i}$ attributed to that state is given by
 \begin{eqnarray}
  \label{costi}
C_{i} = N_{ac}(i) + \beta \cdot \delta_{r}(i).
\end{eqnarray}

As $\beta$ increases, the optimal policy favors maximizing the runway utilization rate over minimizing the number of taxiing aircraft. $\beta$ and $C_{i}$ are expressed in number of aircraft per minute. $\beta$ is the ratio of the cost of non-utilization of the runway for one minute over the cost of having one aircraft on the taxiway for one minute.\\
For each value of $\beta$, the corresponding optimal policy is Pareto optimal and captures the trade-off between minimizing taxi time
 and maximizing runway utilization rate. 
 
\subsubsection{Fairness Considerations when Multiple Ramps are Present}
It is assumed that each terminal (ramp) has aircraft ready to enter the taxiway system, and that they must be served fairly. Two mechanisms have been introduced for that purpose: In {\em ramp alternation}, the policy must serve each ramp once at a time. An additional state is introduced in the Markov Decision process to reflect this. In {\em statistical fairness}, a constraint is introduced to constrain each ramp to be served an equal number of times on average.

\subsection{Information Valuation}
The metrics used to value information are runway utilization and number of taxiing aircraft. 
The value of added information is computed as the improvement in closed-loop system performance generated by this added information.

\subsubsection{Full State Feedback and Optimal Policies}\label{s:LP}
Under full state feedback, the agent controlling the clearances can fully observe the state of the airport surface. The optimal decision $k$ is a function of the observed state $i$. Given the cost structure and the representation of the airport taxi-out process as a Markov Decision Process, it is possible to use linear optimization techniques to find the steady state optimal decision policy $\Pi$ that minimizes the expected cost per time step~\cite{Hiller2001}. If $i(t)$ is the state at time $t$, then
 \begin{eqnarray}
 \mbox{Expected Cost} = \displaystyle\lim_{n \to \infty} E \left( \frac{1}{n} \cdot \displaystyle\sum_{t=0}^n C_{i(t)} \right).
\end{eqnarray}
To detail the optimal control approach, we use the following notations:
\begin{itemize}
\item Let $\iota$ be the state at time $n$.
\item Let $\eta$  be the state at time $n+1$.
\item Let $\kappa$ be the decision variable value at time $n$.
\item Let $y_{ik} = P(\iota=i,\kappa=k)$ be the probability of being in state $i$ and taking decision $k$. The optimal decision $k$ is given by the optimal policy: $k = \Pi(i)$.
\item Let $p_{j|ik} = P(\eta=j|\iota=i,\kappa=k)$ be the probability of transition to the next state $j$ knowing the current state is $i$ and the decision chosen is $k$.
\item In addition, a state is added that describes whether the next pushback originates from ramp 1 or ramp 2 
\end{itemize}

For a steady state process with $M+1$ states and $K$ decisions, the expected cost per time step is \cite{Hiller2001}
   \begin{eqnarray}\label{e:costEquation}
   \displaystyle\lim_{n \to \infty} E(\frac{1}{n} \cdot \displaystyle\sum_{t=0}^n C_{i(t)}) = \displaystyle\sum_{i=0}^M \displaystyle\sum_{k=1}^K C_{i}\cdot y_{ik}.
    \end{eqnarray}

Consequently, the cost function for this linear optimization is
 \begin{eqnarray}\label{e:costEquationFinal}
\mbox{Minimize } Z = \displaystyle\sum_{i=0}^M \displaystyle\sum_{k=1}^K
C_{i}\cdot y_{ik}.
 \end{eqnarray}

Subject to:
\begin{enumerate}
\item Constraints on state-decision probability variables:
\begin{eqnarray} \label{e:stchastiVariableConstraints}
\displaystyle\sum_{i=0}^M \displaystyle\sum_{k=1}^K y_{ik} = 1  \end{eqnarray}
 \begin{eqnarray} \label{e:stochasticVariableConstraints}
  y_{ik} \geq  0, \mbox{for } i = 0..M; k = 1..K \end{eqnarray}

\item Constraints governing state transitions:
 \begin{eqnarray}\label{e:stateTransitionEq}
 \displaystyle\sum_{k=1}^K y_{jk} - \displaystyle\sum_{i=0}^M \displaystyle\sum_{k=1}^K y_{ik}\cdot p_{j|ik} = 0,
\end{eqnarray} \\
\mbox{for } $j = 0..M$; $k =1..K$
\end{enumerate}

Once the optimal set of steady state probabilities of being in state $i$ and taking decision $k$, $y_{ik}$, is evaluated and the corresponding optimal pushback policy is given by

\begin{equation}
\label{e:policyEvaluation}
\Pi_{i} = k \mbox{ with the probability } \frac{y_{ik}}{\sum_k y_{ik}}.
\end{equation}

\subsubsection{Partial Information: Estimated State Feedback} \label{partial}
In this scenario, the agent can observe only a portion of the taxiway system, and he knows the number of taxiing aircraft on the unobservable portion of the taxiway system. Often partial information arises from the fact that the information available in digital form is less complete than that available to human controllers. For example, we have partial information when we design an algorithm based on ACARS/ASPM data only 
Under limited aircraft position information, the system becomes a Partially Observable Markov Decision Process (POMDP). There exists several methods to solve POMDPs optimally \cite{Cassandra1996,Cassandra1997,Sondik1971,Kaelbling1998}. However, these methods are computationally very demanding for a finite time horizon, and they are not appropriate for an infinite time horizon. Indeed, finite-horizon POMDPs are PSPACE-complete \cite{Papadimitriou1987} and infinite-horizon POMDPs are undecidable \cite{Madani1999}.

\begin{figure}[!t]
\centering
\includegraphics[scale = .4]{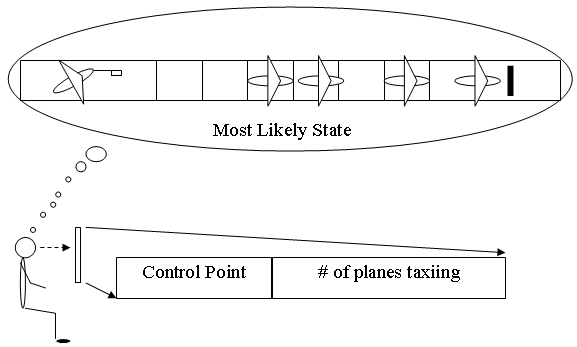}
\caption{Estimation of the Taxiway System State by a Decision Maker}
\label{MostLikelyState}
\end{figure}

\paragraph{Most Likely State}
For these reasons, methods applicable to an infinite time horizon and computationally more tractable were considered \cite{Cassandra1996}. The Markov process that is modeled for LaGuardia includes more than 220,000 transitions with non-zero probabilities. Heuristic methods are computationally faster, and they are better suited to determine effective control laws for this POMDP. Among these, the Most Likely State (MLS) algorithm was selected because it is applicable to an infinite time horizon and compares favorably with other heuristic algorithms \cite{Cassandra1996}. Moreover, its steps resemble the behavior of a decision maker under uncertainty. Indeed, this heuristic control strategy consists of estimating the most likely current state, and choosing the corresponding optimal decision, using the optimal decision policy evaluated in the full state feedback case. However, the decision is not optimal as will be seen later. Fig.~\ref{MostLikelyState} illustrates the information available to the decision maker.

The variables used in the MLS algorithm are
\begin{itemize}
\item $\Theta$, the index of the current observation,
\item $b_i$, the probability of the airport model to be in state $i$ for all states $i$ of the state space, ($(b_i)_{i \in I}$ is the belief state vector.)
\item and $p_{o|j} = P(\Theta=o|\eta=j)$, the probability of observing $o$, knowing the current state is $j$, for all observations $o$ in the observation space, and for all states $j$ in the state space.
\end{itemize}
The updater function takes the previous belief state $b$, the current observation $o$, the previous decision $k$, and returns the new belief state vector  $b'$.\\ The following equation aims at updating belief states and is derived from Bayes' rules \cite{Littman1994}:
\begin{equation}
  \label{e:updater}
 b'_{j} = \frac{p_{o|j} \displaystyle\sum_{i=0}^M p_{j|ik} \cdot b_{i} }{ \displaystyle\sum_{j=0}^M p_{o|j} \displaystyle\sum_{i=0}^M p_{j|ik} \cdot b_{i}}.
 \end{equation}

\begin{figure}[!t]
\centering
\includegraphics[width=3.5in]{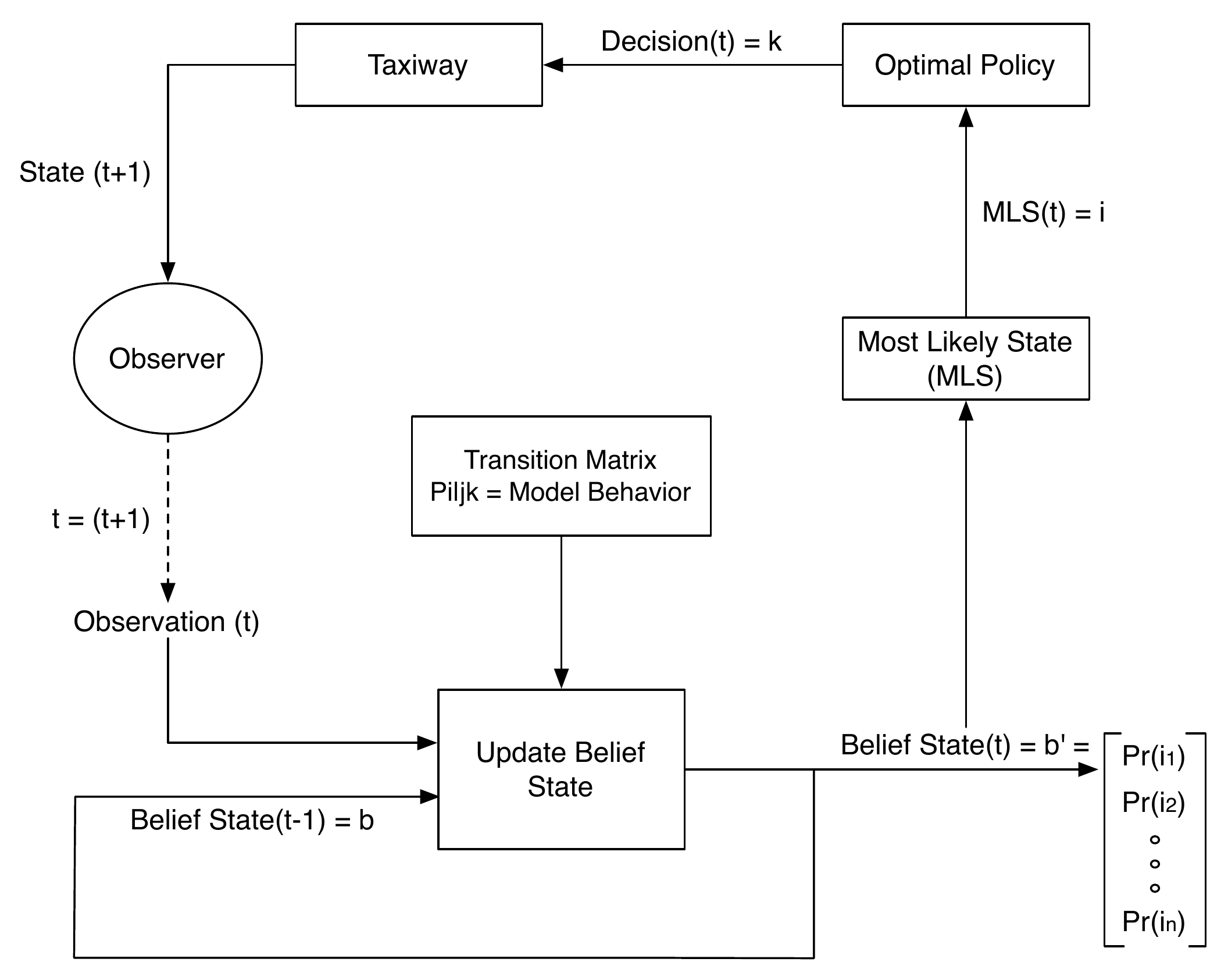}
\caption{Heuristic Control of Taxi Clearance Decisions Based on Partial Observation}
\label{estimator}
\end{figure}

Fig.~\ref{estimator} details the heuristic control of taxi clearance decisions based on partial observations.

\paragraph{Observation Probability Matrix}
The information contained in an observation is given by the probability matrix $p_{o|j}$. This probability is key to evaluate the probability of having every state $j$, given a specific observation $o$ and previous belief $b$. Eq. (\ref{e:updater}) explains how observations of the surface are incorporated into the decision process during the update of the belief state.

Assume that the agent can observe only the control point and he knows the number of taxiing aircraft. Let $O$ be the observation space and $c$ be the total number of components, or piece of information, included in each observation $o \in O$. Then $O$ is a subset of $\Re^c$. In this scenario, there are two pieces of information, $c = 2$, the number of taxiing aircraft $N_{ac}$, and whether or not it is physically possible to clear an aircraft using ($RampFree$), a binary variable.
An observation is a vector defined by

\begin{equation} \label{e:oVect0}
o = \begin{bmatrix} N_{ac} & RampFree \end{bmatrix}.
\end{equation}

The algorithm generating the observation probability matrix uses an injective function,  which attributes a unique observation index $n(o)$ to every observation $o$. The injective function converts the observation vector with 2 components into a binary vector of $[\text{roundup}(\text{log2}(\text{max}(Nm_{ac})))+\text{roundup}(\text{log2}(\text{max}(RampFree)))]$ bits to then reconvert it back to its decimal value, as illustrated in Eq. (\ref{e:obsInjFunct}). \text{roundup} function gives the closest integer of the argument.

\begin{equation}
\label{e:obsInjFunct}
n(o) = \text{bin2dec}(\begin{bmatrix} \text{dec2bin}(N_{ac}) & \text{dec2bin}(RampFree)\end{bmatrix})
\end{equation}

For any state $j$, there exists only one information that can be observed $o(j)$, consequently for a system with $N$ possible observations and $M$ states, observation probabilities are zeros and ones, i.e. $\forall (o_n,j) \in \{1..N\}\times \{1..M\}$, $p_{o_n|j} \in \{0,1\}$. Eq. (\ref{e:obsPoj}) shows how the $p_{o|j}$ matrix is evaluated.

\begin{equation}\label{e:obsPoj}
p_{o_n|j} =
\begin{cases}
1 \text{ if $o_n = n(o(j))$}, \\
0 \text{ if $o_n \neq n(o(j))$}.
\end{cases}
\end{equation}

\subsubsection{Threshold Policy}
A threshold policy is a pushback control law, which relies solely on the current number of taxiing aircraft to make a push-back decision as described in~\cite{log99,SKB:11}. This simple control law computes the number of taxiing aircraft  $N(i)$ for state $i$ and compares it to a given threshold value $Th$ \cite{log99}. If the number of aircraft is greater than the threshold, no pushback clearance is issued, and $k=0$. On the other hand, if that number is smaller than the threshold, a pushback clearance is issued, and $k=1$. This is summarized by 
\begin{equation} \label{e:theq1}
k =
\begin{cases}
0 \text{  if $N(i) > Th$},\\
1 \text{  if $N(i) \leq Th$}.
\end{cases}
\end{equation}

Note that the threshold policy can be evaluated analytically since the corresponding closed-loop system is a Markov chain. When multiple ramps are present, the threshold policy is required to alternate evenly among the ramps.

\subsection{Optimal Policies Against Benchmark Policy: New York LaGuardia Airport}
\begin{figure}[!t]
\centering
\includegraphics[width=3.5in]{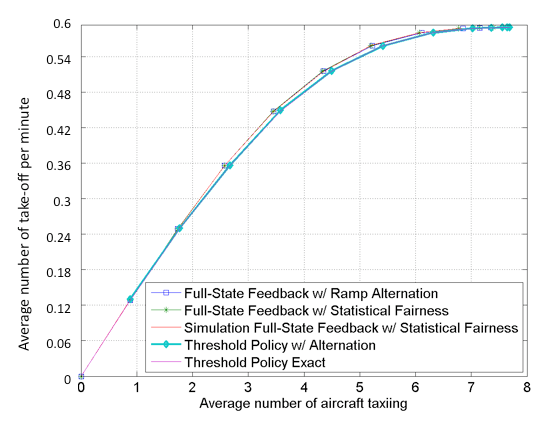}
\caption{Average Take-off Rate as a Function of the Average Number of Taxiing Aircraft at LaGuardia Airport}
\label{f:CXutilVsTax}
\end{figure}

\begin{figure}[!t]
\centering
\includegraphics[width=3.5in]{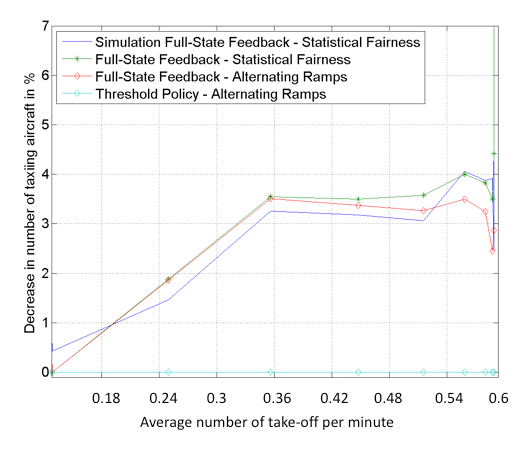}
\caption{Reduction in Percent of the Average Number of Taxiing Aircraft as a Function of the Take-off Rate, when Compared with a Threshold Policy which Alternates Between Ramp One and Ramp Two.}
\label{f:CximprovVsUtil}
\end{figure}

Fig.~\ref{f:CXutilVsTax} illustrates the average take-off rate for the two-ramp model of the LaGuardia Airport, as a function of the average number of taxiing aircraft. It is shown that there is no significant difference between full-state feedback with ramp alternation and full-state feedback with statistical fairness. The statistical fairness does not mean to release departures from alternating ramp areas every time. It just guarantees the overall release from each ramp area is balanced on average. Also, the performance of the threshold policy is similar to that of the full-state feedback. Fig.~\ref{f:CximprovVsUtil} shows the reduction in percent of the average number of taxiing aircraft for optimal policies, as a function of the take-off rate, when compared with a threshold policy that releases aircraft from alternating ramp areas. Hence, Fig.~\ref{f:CximprovVsUtil} shows the impact of full-state feedback on the number of taxiing aircraft, or equivalently congestion on the airport surface, while maintaining the runway throughput. 

When the number of taxiing aircraft is limited to one aircraft by the threshold policy, the difference of performance between the full-state feedback policy and the threshold policy is non-existent, as illustrated by Fig.~\ref{f:CximprovVsUtil}. This confirms the intuition that there is no benefit in knowing the exact position of aircraft when there can be no conflict between aircraft on the taxiway. However, this case is not realistic in practice. 

When the threshold for the number of taxiing aircraft is increased to two and and then three, which corresponds to the take-off rate of 0.0.27 and 0.4, the threshold policy starts yielding a lower take-off rate for the same number of taxiing aircraft than the full-state feedback policy, as shown in Fig.~\ref{f:CXutilVsTax} and \ref{f:CximprovVsUtil}. Indeed, the threshold policy releases aircraft blindly, based on the number of taxiing aircraft. So, the threshold policy does not recognize which part of taxiway system is congested with departures. For instance, when all the taxiing aircraft are packed between ramp 1 and ramp 2, releasing a flight from ramp 2 would help prevent the runway from being starved. However, the threshold policy would not release more aircraft until the number of taxiing aircraft becomes below the threshold, and it will result in reducing the utilization rate of the runway. On the other hand, the optimal full-state feedback policy performs better because it manages the release of aircraft using the exact position of the other aircraft already taxiing.

Among the two fairness rules described above, the policy based on statistical fairness yields the best results, as shown in Fig.~\ref{f:CximprovVsUtil}. However, it performs close to the policy that strictly alternates between ramps. It is noticeable that the simulation of the statistical fairness optimal policy, produces performances that are slightly worse than those directly indicated by the optimization software output.

As shown in Fig.~\ref{f:CximprovVsUtil}, the full state feedback policies perform consistently better, generating a smaller average number of taxiing aircraft, when compared with the threshold benchmark policy. This performance is consistently better over a wide array of runway utilization rates, which correspond not only to intermediate runway capacities, but also to situations where the runway is used at maximum capacity. For take-off rates beyond 0.36, the reduction of the number of taxiing aircraft is consistently above 3.5 percent.

\subsection{Influence of Different Levels of Observation: Seattle Tacoma Airport}
\begin{figure}[!t]
\centering
\includegraphics[width=3.5in]{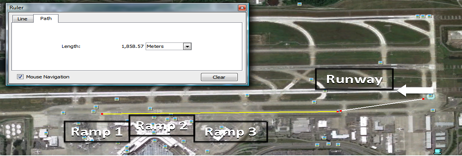}
\caption{Seattle Airport in Most Common Configuration with Departures on Runway 16L.}
\label{f:SEA}
\end{figure}

The methodology presented here has been extended to other airports with simple runway/taxiway structures. Seattle Tacoma airport is modeled with three ramp areas as shown in Fig.~\ref{f:SEA}. Similarly to the LaGuardia airport model, each spatial sample represents 200 meters. In this model, the taxiway stochastic properties are chosen consistently with the previous calibrations of LaGuardia airport model. Two Bernoulli variables with parameters $c_1$ and $c_2$ are are set to reproduce the standard deviation (0.603 aircraft/minute) and average (0.712 aircraft/minute) take-off rate that Seattle airport reaches at saturation. 

\begin{figure}[!t]
\centering
\includegraphics[width=3.5in]{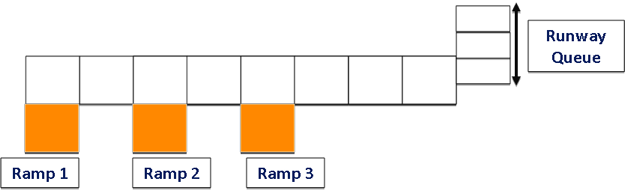}
\caption{Partial Observation Level One, at Seattle airport, Observations Include the Control Points and the Total Number Taxiing Aircraft.}
\label{f:level1}
\end{figure}

\begin{figure}[!t]
\centering
\includegraphics[width=3.5in]{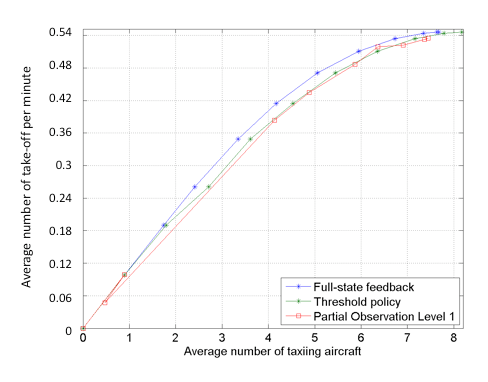}
\caption{Average Take-off Rate as a Function of the Average Number of Taxiing Aircraft at Seattle Airport for Full-State Feedback, Threshold Policy, and Partial Observation Level One.}
\label{f:level1result}
\end{figure}

Two levels of partial information are studied to capture the performance differences for intermediate levels. The first level of information includes the total number of taxiing aircraft, and indicates if taxi clearance from each ramp area is feasible, which is specified by whether each orange spatial sample in Fig.~\ref{f:level1} is occupied by an aircraft or not. This level of information is easily available without any surveillance system because the spot and the nearby taxiway are visually observable by ramp controllers. The observation vector corresponding to the level one of information is defined by 
\begin{equation}
o = \begin{bmatrix} N_{white} & N_{ramp1} & N_{ramp2} & N_{ramp3} \end{bmatrix}.
\end{equation}
The MLS algorithm takes this observation vector in order to estimate the most likely state and get the optimal decision given the estimated state. Fig.~\ref{f:level1result} illustrates the take-off rate for the Seattle Airport model. The optimal policy with the full state feedback decreases the average number of taxiing aircraft by 6\% while maintaining a take-off rate above 0.54, compared with the threshold policy. The first level of information produces take-off rates slightly lower than the threshold policy does at the same level of congestion on the taxiway system, which is specified by the number of taxiing aircraft. The estimated state feedback policy estimates the most likely state given the partial observation and makes the optimal decision based on the most likely state. The partial observation level one contains so little information that the most likely state is significantly far from the actual state. Therefore, the partial observation level one leads to worse decision than the threshold policy does although the partial observation level one has more information than the threshold policy. 

\begin{figure}[!t]
\centering
\includegraphics[width=3.5in]{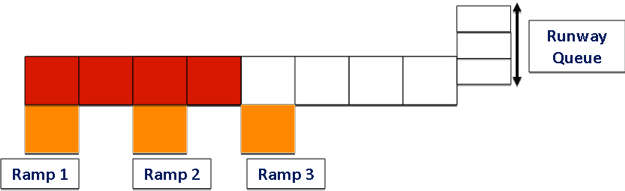}
\caption{Partial Observation Level Two, at Seattle airport, Surface Surveillance Covers the Taxiway System in Front of the Ramps.}
\label{f:level2}
\end{figure}

\begin{figure}[!t]
\centering
\includegraphics[width=3.5in]{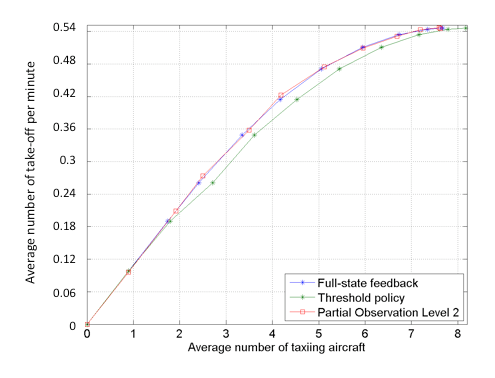}
\caption{Average Take-off Rate as a Function of the Average Number of Taxiing Aircraft at Seattle Airport for Full-State Feedback, Threshold Policy, and Partial Observation Level Two.}
\label{f:level2result}
\end{figure}

The second level of information is illustrated in Fig.~\ref{f:level2}. The agent can observe the taxiway system in front of the ramps. In real-life situation, this level of partial information is also relatively easy to be obtained without the help of surveillance system. Ramp controllers are able to observe the portion of the taxiway near the corresponding ramp area visually, and the position of aircraft can be determined with respect to each ramp. For instance, ramp controllers can tell an aircraft is located by the ramp or between two ramps. Because the length of a spatial sample is 200 meters and the ramp is like a landmark, such a rough observation is sufficient to tell the location of an aircraft near the ramp. The observation vector corresponding to the second level of information is defined by 
\begin{equation}
o = \begin{bmatrix} & N_{white} & N_{red1} & N_{red2} & N_{red3} & N_{red4} & N_{ramp1} \nonumber \\
 & N_{ramp2} & N_{ramp3} \end{bmatrix}.
\end{equation}
As more information is provided to the agent or a computer aid, the accuracy and efficiency of the MLS algorithm increases. Therefore, the difference between the most likely state and the actual state becomes smaller than those from the partial observation level one, and the MLS algorithm performs close to the Full state feedback policy as illustrated in Fig.~\ref{f:level2result}.

The study of different levels of observation for Seattle airport suggests that the utility of surveillance information for airport departure operations increases significantly when it directly helps manage conflicts on the airport surface.
\section{Conclusions}
This paper assesses the benefits of providing surface surveillance information to the ramp clearance control process at busy airports.

Our results have shown that, within a collaborative framework allowing the creation of a virtual queue, surface surveillance information can significantly improve the control of stochastic departure operations on the ground. More specifically, at LaGuardia airport, controlling taxi clearances optimally using surface surveillance reduces the number of taxiing aircraft by 4\% and therefore emissions by as much when the airport operates near capacity, compared with a threshold policy which limits the number of taxiing aircraft. At Seattle airport, controlling taxi clearances optimally using surface surveillance reduces the number of taxiing aircraft by 6\% when the airport operates near capacity, compared with a threshold policy which limits the number of taxiing aircraft. It has been observed that, in order to minimize wasteful surface conflicts and queues, the optimal full-state feedback policy relies on aircraft position information to avoid conflicts, maximize runway utilization, and balance and coordinate ramp taxi clearances. However, the spot-release strategies presented in this paper may induce congestion on the ramp by holding departures at spots, and taxi times and emissions remain because the aircraft held at the spot are still taxiing with engines on. In order to maximize the operational and environmental benefits from the spot-release strategies, the control position needs to be shifted toward the gates and further analysis on the impact of holding aircraft at the gates on the ramp operations is necessary. For example, holding departures at gates may cause gate shortage, which results in additional delays and costs. In addition, future study would extend the airport surface model to the gates with the same methodology. Currently, ASDE-X does not cover the ramp area, but the movements of aircraft on the ramp area are as important as on the AMA. Also, a potential future work would analyze partial information on the runway queue rather than near the ramps because it was shown that the way partial information has been looked at (close to ramps) in this paper is not very smart.

\appendices

\bibliographystyle{IEEEtran}
\bibliography{IEEEabrv,293burga} 


\begin{IEEEbiographynophoto}{Pierrick Burgain}
received the M.S. degree from \'Ecole Sup\'erieure d'Electricit\'e, Gif-sur-Yvette, France, and the M.B.A. degree and the Ph.D. degree in electrical and computer engineering from the Georgia Institute of Technology, Atlanta. He was an Analyst with Metron Aviation. He is currently a Business Manager with Capital One, Atlanta. His research interests include stochastic control and optimization and their applications in air traffic operations.
\end{IEEEbiographynophoto}

\begin{IEEEbiographynophoto}{Sang Hyun Kim}
is a Ph.D. candidate in the School of Aerospace Engineering at the Georgia Institute of Technology. He holds his B.S. degree in mechanical and aerospace engineering from Seoul National University, South Korea. Sang Hyun Kim’s research interests are optimization, transportation, airport operations, ramp management, and general aerospace engineering. He currently works on the project of optimizing gate assignments of Delta Air Lines in Hartsfield-Jackson Atlanta International Airport.
\end{IEEEbiographynophoto}

\begin{IEEEbiographynophoto}{Eric Feron}
received the B.S. degree in applied mathematics from Ecole Polytechnique, Palaiseau, France, the M.S. degree in computer science from the Ecole Normale Sup\'erieure, Paris, France, and the Ph.D. degree in Aeronautics and Astronautics from Stanford University, Stanford, CA. He is the Dutton-Ducoffe Professor of Aerospace Software Engineering, Georgia Institute of Technology, Atlanta. He is also a consulting professor with the Ecole Nationale de l'Aviation Civile, Toulouse, France. Prior to that, he was with the faculty of the Department of Aeronautics and Astronautics, Massachusetts Institute of Technology, Cambridge, for 12 years. His former research students are distributed throughout academia, government, and industry. He has published two books and several research papers. His research interests include using fundamental concepts of control systems, optimization, and computer science to address important problems in aerospace engineering such as aerobatic control of unmanned aerial vehicles, multiagent operations such as air traffic control systems, and aerospace software system certification.
\end{IEEEbiographynophoto}

\end{document}